\documentclass[aps,prl,twocolumn,showpacs,preprintnumbers,amsmath,amssymb,superscriptaddress]{revtex4-1}
\usepackage{graphicx}
\DeclareMathOperator{\sech}{sech}

\begin{document}

\title{Coherent inverse Compton scattering with attosecond electron bunches accelerated and compressed by radially polarized laser pulses}

\author{A. Sell}
\email[]{asell@mit.edu}
\affiliation{Department of Electrical Engineering and Computer Science and Research Laboratory of Electronics, Massachusetts Institute of Technology, 77 Massachusetts Ave., Cambridge, MA-02139, USA}
\author{F. X. K\"artner}
\affiliation{Department of Electrical Engineering and Computer Science and Research Laboratory of Electronics, Massachusetts Institute of Technology, 77 Massachusetts Ave., Cambridge, MA-02139, USA}
\affiliation{Center for Free-Electron Laser Science, Deutsches Elektronen Synchrotron, and Department of Physics, University of Hamburg, Notkestra\ss e 85, 22607 Hamburg, Germany}
\date{\today}

\begin{abstract}
We present a study of direct laser driven electron acceleration and scaling of attosecond bunch compression in unbound vacuum. Simple analytical expressions and detailed three-dimensional numerical calculations including space charge reveal the conditions for compression to attosecond electron sheets. Intermediate emittance minima suitable for brilliant x-ray generation via coherent inverse Compton scattering (ICS) are predicted. We verify the coherent emission properties of the resulting x-ray fields and demonstrate feasability for realistic laser parameters.
\end{abstract}

\pacs{41.75.Jv, 41.60.Cr}

\maketitle

Ultrashort relativistic electron bunches and brilliant x-ray bursts are expected to open new avenues in ultrafast physics, serving as probes with atomic spatial and attosecond temporal resolution \cite{r09chapman}. Applications like ultrafast molecular chemistry, monitoring structural dynamics of physical and dynamical processes and phase-contrast imaging fuel the ever growing demand for beamtime at synchrotrons and free-electron lasers and drive a quest for laboratory scale sources \cite{r09graves}.

Electron bunches in the order of 10~fs have recently been achieved with plasma-wakefield accelerators \cite{r10debus, r11lundh}. Certain issues of this technique, like instabilities of the nonlinear interactions and a plasma-limited compression rate, are circumvented by vacuum laser acceleration. Few-femtosecond bunches have been predicted for approaches exploiting the ponderomotive force \cite{r01stupakov, r04kong}. However, the efficiency of this indirect interaction is limited, and quiver motions distort the transverse phase space \cite{r07karmakar}.

Recent work on direct laser acceleration of electrons shows promise in overcoming these issues: The axial electric field of radially polarized pulses efficiently accelerates electrons, whereas the radial fields provide transverse confinement \cite{r04birula}. Extensive work on single electrons travelling on-axis has shown the feasibility of the acceleration concept \cite{r06salamin, r10fortin, r10wong}. Three-dimensional simulations have predicted attosecond pulse durations, however without considering space-charge and bunch emittance effects or by concentrating on PW-class driving lasers \cite{r06varin, r07salamina, r07karmakar}.

Here, we propose a general scheme to generate low-emittance attosecond electron bunches to produce brilliant coherent x-ray bursts in a laser undulator. Our example yielding 15~as bunches at 8.8~MeV assumes a moderate driving laser power of only 10~TW, which might soon be available even at kHz repetition rates. We give a physical understanding and scaling of the bunch compression, and investigate the influence of space charge, transversal dynamics and particle statistics. We find intermediate acceleration phases exposing extremely low emittance values at high electron energies. Our numerical simulations use a relativistic three-dimensional model, including the fields emitted by the electrons. Assuming a combination of a radially polarized pulse for the compression and acceleration and a counterpropagating linear polarized Gaussian pulse for the wiggling of the electrons, we discuss the coherent emission properties of the resulting x-ray pulses.

Bunch compression very naturally arises in direct laser acceleration in unbound vacuum: Assuming a fixed time for each electron to get accelerated from an initial velocity $\beta_i (= v_i / c)$ to the final velocity $\beta_f$, a field travelling with the speed of light $c$ takes the time $T_{\rm cpt} = L_i c^{-1} / \left(1-\beta_i\right)$ to transverse a bunch of length $L_i$. The final bunch extension is given by the velocity mismatch between field and electrons \cite{r07wang}
\begin{equation}
	\label{eq02}
	L_f = \left(\beta_\phi-\beta_f\right) c T_{\rm cpt} = \frac{\beta_\phi-\beta_f}{1-\beta_i} L_i \approx \frac{1}{2 \gamma_f^2} \frac{1}{1-\beta_i} L_i,
\end{equation}
with the approximation $\gamma_f \gg 1$ (with $\gamma = 1/\sqrt{1-\beta^2}$) and $\beta_\phi \equiv 1$, for now. For $\beta_f > \beta_i$ the bunch obviously compresses, even in its rest frame where there is no relativistic contraction. From Eq.~\ref{eq02} we obtain a temporal compression factor $\eta$ of
\begin{equation}
	\label{eq03}
	\eta = \frac{\beta_i L_f}{\beta_f L_i} = \frac{1-\beta_\phi \beta_f^{-1}}{1-\beta_i^{-1}} \approx \frac{1}{2 \gamma_f^2} \frac{1}{\beta_i^{-1}-1}.
\end{equation}
Note that the wavelength achieved by ICS also scales as $1 / \gamma_f^2$ and stays matched to the bunch length while scaling the energy of the proposed coherent scheme. Compression gets detoriated mainly by space charge, radial field inhomogenities, and the initial electron momentum distribution. In the following, we identify situations where $L_f$ is minimized.

Our radially polarized acceleration laser is described with Gaussian parameters ($E_0$: peak field, $w_0$: waist radius, $w = w_0 \sqrt{1+ z^2/z_R^2}$: beam radius, $\omega$: angular frequency, $z_R = \omega w_0^2 / \left(2c\right)$: Rayleigh length, $\tau = t-z/c$: retarded time, $t_0$: pulse duration, $\Psi_0$: phase). In cylindrical coordinates $(r, \phi, z)$ the nonzero components are given by
\begin{eqnarray}
	\label{eq04}
	\Psi &=& \Psi_0 + \omega \tau - \frac{z}{z_R} \frac{r^2}{w^2} + 2 \arctan{\left( \frac{z}{z_R} \right)}\\
	\nonumber
	{\mathcal E} &=& E_0 \frac{w_0^2}{z_R w^2} \sech{\left(\frac{\tau}{t_0}\right)} e^{-\frac{r^2}{w^2}} \\
	\label{eq05}
	E^{\rm ex}_r &=& cB^{\rm ex}_\phi = {\mathcal E} r \cos{\left(\Psi\right)}\\
	\label{eq06}
	E^{\rm ex}_z &=& {\mathcal E} \frac{w_0^2}{z_R}
	\left[\left( 1-\frac{r^2}{w^2} \right) \sin{\left(\Psi\right)}
	- \frac{z}{z_R}\frac{r^2}{w^2} \cos{\left(\Psi\right)} \right].
\end{eqnarray}
On the center axis ($r=0$) the radial field $E^{\rm ex}_r$ vanishes, and $E^{\rm ex}_z$ leads to direct acceleration. The on-axis phase velocity is derived from Eq.~\ref{eq04} with $d\Psi = {\rm const.}$, and exceeds $c$ due to the Gouy phase:
\begin{equation}
	\label{eq07}
	\beta_\phi = \left[ 1 - \frac{4 c^2}{\omega^2 w^2} \right]^{-1} > 1.
\end{equation}
The electrons travel with $\beta_{\rm el.} < 1$ and thus slip through acceleration periods of the accelerating field. We estimate the phase slip time $T_{\rm slip}$ assuming a mainly constant electron energy during one acceleration-cycle by 
\begin{equation}
	\label{eq08}
	T_{\rm slip} \left( \beta_\phi - \beta_{\rm el.} \right) = 2 \pi / \omega.
\end{equation}

The equations of motion for each particle are
\begin{eqnarray}
	\nonumber
	{\rm d}_{c t} \left( \vec \beta \gamma \right) &=& \frac{q}{m_0 c^2} \left[ \left(\vec E^{\rm ex} + \vec E^{\rm in}\right) + \vec \beta \times c\left( \vec B^{\rm ex} + \vec B^{\rm in} \right) \right]\\
	\label{eq09}
	{\rm d}_{c t} \vec R &=& \vec \beta = \frac{\vec \beta \gamma}{\sqrt{1 + \left( \beta \gamma \right)^2}}.
\end{eqnarray}
Many studies solve equations for ${\rm d}_t \vec \beta$ instead of Eq.~\ref{eq09}, causing numerical problems for $\beta \to 1$. The internal fields $\vec E^{\rm in}_n$ and $\vec B^{\rm in}_n$, acting on particle $n$, are calculated by summing the Lorentz-transforms of the Coulomb fields evaluated in the rest frames of all particles $m \ne n$.

We consider an accelerator geometry, where electron bunches with an energy of $W_0 = 100~{\rm keV}$ (i.e. from a DC electron gun) get injected. This nonrelativistic energy yields similar results as acceleration from rest, but avoids issues like creating a bunch in the focus and excessive space charge effects \cite{r07salamina, r07karmakar}. We assume a ellipsoidal bunch, both in real and momentum space. All diameters and energy spreads are set to 2~$\mu$m and $\Delta W / W_0 = 0.01$, respectively, in order to account for a strong focussing. These parameters are very challenging, only recently durations shorter than the 12~fs assumed here have been demonstrated by field emission from nano-tips \cite{r06hommelhoff}. Appropriate charges might be obtained from arrays of tips. An injection parameter $z_0$ defines the time $ct = - z_0 / \beta_i$, at which the bunch would reach the origin $z=0$ without external fields. Unless mentioned otherwise, a moderate 10~TW laser with $w_0 = 2 \, \mu {\rm m}$, $\lambda = 2 \pi c / \omega = 800 \, {\rm nm}$, a FWHM pulse duration of 10~fs ($t_0 = 5.7 \, {\rm fs}$) and an energy of 100~mJ is assumed in the rest of the manuscript.

\begin{figure}
	\includegraphics[width = 8.3 cm]{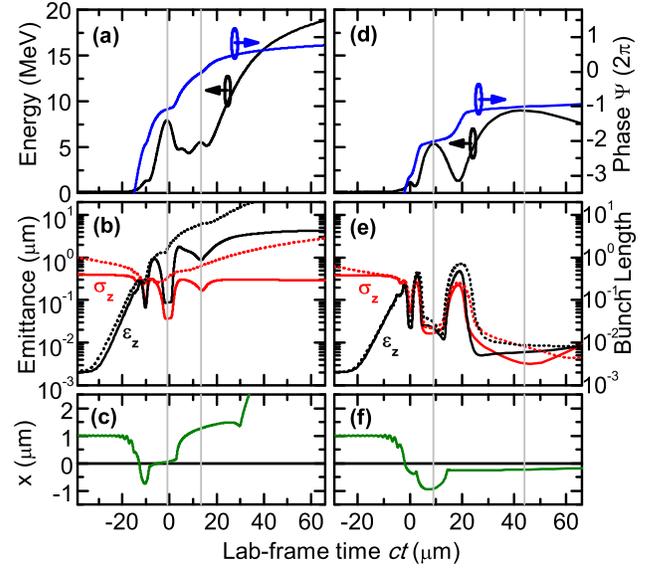}
	\caption{Evolution of (a) center bunch energy (black) and on-axis phase $\Psi$ (blue), (b) bunch emittance $\epsilon_z$ (black) and length $\sigma_z$ (red), and (c) transversal position $x$ of one particle at the bunch equator during acceleration from 100 keV to 19~MeV. Laser phase $\Psi_0$ and injection offset $z_0$ are optimized for final electron energy. (d-f) Acceleration from the same starting parameters optimized for minimum intermediate emittance. Dotted lines in (b,e) are calculated for a bunch charge of $Q = - 200~{\rm fCb}$, solid lines without space charge. Vertical gray lines indicate acceleration phases between slippage events.\label{fig01}}
\end{figure}

Fig.~\ref{fig01}~(a) shows the average electron energy and the central laser phase $\Psi$ as evolving with time $ct$. We set $z_0 = - 3.7 \, \mu {\rm m}$ and $\Psi_0 = 5.26$ to achieve a maximum final average bunch energy of 19~MeV \cite{r10fortin, r10wong}. Vertical grey lines mark maxima in energy, corresponding to $2 \pi$-multiples of the phase. The main acceleration takes place while the phase approaches $\Psi = 2 \pi$. Assuming a slippage time of $c T_{\rm slip} = 17 \, \mu{\rm m}$ and highest phase velocity, Eqs.~\ref{eq07} and \ref{eq08} yield a longitudinal electron energy of only $1.6$~MeV, indicating strong transversal distortions of the bunch during acceleration.

Fig.~\ref{fig01}~(b) depicts the normalized rms-Emittance $\epsilon_z$ and the rms bunch length $\sigma_z = \sqrt{ \left< z^2 \right> - \left< z \right>^2 }$. We use the standard definition $\epsilon_u = \frac{1}{m_0 c} \sqrt{ \left< u^2 \right>_c\left< p_u^2 \right>_c - \left< u p_u \right>_c^2}$ where $\left< f g \right>_c = \left< f g \right> - \left< f \right>\left< g \right>$ is a centered average over all particles and $(u, p_u)$ with $u \in \{ x, y, z\}$ denote cartesian coordinates and momenta. For relativistic energy spreads the rms-Emittance generally is not conserved(!), even without space charge and external fields. This is because $z$ changes with $\beta_z$, which is a nonlinear function of $p_z = m_0 c \gamma \beta_z$. Thus, the phase volume deforms, and $\epsilon$ changes without violating Liouville's theorem.

The 5000-fold increase of emittance in Fig.~\ref{fig01}~(b) appears mainly in the region close to $ct = 0$, where slippage occurs fast due to the Gouy-phase and the transversal dynamics. $\sigma_z$ stays nearly unaltered, which we attribute to the compression expected from Eq.~\ref{eq02}. A transversal collapse occuring close to $ct = 0$ and $z=0$ (Fig.~\ref{fig01}~(c)) makes the process highly sensitive to space charge, as seen from the dotted lines in (b). Together with the strong emittance-growth, this effect severely affects the usefulness of this scheme for experiments.

Both issues are solved by choosing a different laser phase $\Psi_0 = 4.87$ and injection timing $z_0 = + 1.9 \, \mu {\rm m}$ (Fig.~\ref{fig01}~(d-f)). The main acceleration now occurs at $\Psi = - 2 \pi$ and the electrons pass the focus before the peak field arrives. A slippage time of $T_{\rm slip} = 35 \, \mu{\rm m}$ (see grey lines) corresponds well to an average electron energy of $3.9 \, {\rm MeV}$. In comparison to (a), the final energy is smaller by a factor of 4, but whenever $\Psi$ is close to a multiple of $2 \pi$, the energy peaks and $\epsilon_z$ and $\sigma_z$ go through pronounced minima which can be exploited for coherent x-ray generation. The highest peak at $ct = 45 \, \mu {\rm m}$ exhibits an energy of 8.8~MeV and a normalized longitudinal emittance of 5~nm, which amounts to only 2.5 times the initial value. The bunch duration is compressed by a factor of $\eta = 200$ to a value as low as 10~as, in reasonable agreement with $\eta = 520$ calculated from Eq.~\ref{eq03}. The typical extension of the low-emittance regions is in the order of $T_{\rm slip} / 4$. Including a space charge of 0.2~pCb, $\epsilon_z$ and $\sigma_z$ only increase by a small factor of 2 (dotted lines in Fig.~\ref{fig01}~(e)).

\begin{figure}
	\includegraphics[width = 8.3 cm]{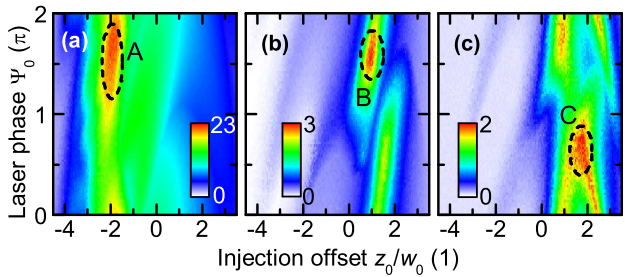}
	\caption{Optimization of acceleration laser phase offset $\Psi_0$ and injection offset $z_0$ for (a) peak electron energy (MeV), (b) peak of electron energy over bunch length (MeV/nm) and (c) peak of electron energy over emittance (MeV/nm). Parameters: $N=400$, no space charge, 100 keV average injection energy.\label{fig02}}
\end{figure}
In Fig.~\ref{fig02} we show the optimization of $\Psi_0$ and $z_0$ to select the values used above. Each point in these plots shows a figure of merit (FOM) evaluated in one simulation run. To speed up calculation, we use only 400 particles and no space charge. The low particle count does not affect the results noticably, as we checked at several test points.
Fig.~\ref{fig02}~(a) displays a color coded plot of the peak electron energies. The maximum is located at negative $z_0$ and a laser phase $\Psi_0$ close to zero. Essentially the same results are obtained optimizing for final electron energy. Fig.~\ref{fig02}~(b) shows peak values of the electron energy $W$ divided by the bunch length $\sigma_z$. This FOM yields similar results as optimizing for $1/\sigma_z$, however avoiding situations with too little energy gain. As indicated by the dashed circles A and B, the regions of peak energy and best compression are clearly disjunct. Finally, Fig.~\ref{fig02}~(c) depicts the maxima of energy over $\epsilon_z$. The peak regions obtained in both latter cases overlapp well, so optimizing on either one helps avoiding the excessive distortions of the bunch occuring otherwise.

\begin{figure}
	\includegraphics[width = 8.3 cm]{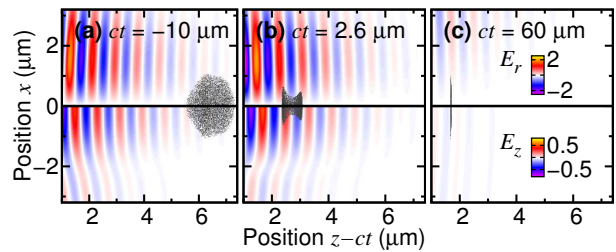}
	\caption{Steps of the acceleration process depicted in the lab-frame. (a) Onset of modulations. (b) Slippage event at an emittance maximum. (c) Acceleration at minimum emittance and bunch duration. Interacting particles are represented by grey dots, normalized transversal (upper panel) and axial (lower panel) electric fields are color coded. Total charge: $Q = - 200~{\rm fCb}$; Particle number: $N = 10000$.\label{fig03}}
\end{figure}
Fig.~\ref{fig03} shows snapshots of the bunch evolution in our scheme. In subfigure~(a), slight modulations of the electron density set in, and the emittance starts increasing. Fig.~(b) shows the bunch while the emittance peaks and the electrons slip from one acceleration phase into the next. Subfigure~(c) sketches the sheet-like bunch while the energy peaks and $\sigma_z$ is minimal. The 15-as-duration of the entire bunch allready includes curvature and also favours the usage the intermediate energy peaks instead of the strongly curved final bunch.

Finally, we investigate brilliant x-ray generation by ICS with the optimized bunches. A counterpropagating linearly polarized Gaussian laser field is superimposed to the acceleration laser. Its non-vanishing cartesian components are given by $\Psi = \Psi_0 + \omega \left(\tau + \zeta/c\right) + \frac{\zeta}{z_R} \frac{r^2}{w^2} - \Psi_{\rm Gouy}$, ${\mathcal E} = E_0 \frac{w_0}{w} \sech{\left(\frac{\tau + \zeta/c}{t_0}\right)} e^{-\frac{r^2}{w^2}}$, $E^{\rm ex}_x = -cB^{\rm ex}_y = {\mathcal E}  \sin{\left(\Psi\right)}$ and $y E^{\rm ex}_z = - x cB^{\rm ex}_z = {\mathcal E} \frac{x y w_0}{w z_R}	\cos{\left( \Psi-\Psi_{\rm Gouy} \right)}$, with $\zeta = z-z_{\rm off}$ and $c\tau = ct-z_{\rm off}$, the Gouy phase $\Psi_{\rm Gouy} = \arctan{\left( \frac{\zeta}{z_R} \right)}$, a wavelength $\lambda = 1.0 \, \mu{\rm m}$, a pulse energy of 20~mJ, a beam radius of $w_0 = 5 \, \mu {\rm m}$, a FWHM duration of 31~fs ($t_0 = 18 \, {\rm fs}$) and all other quantities as defined for Eq.~\ref{eq04}. Despite the electron dynamics, we now also calculate the radiated electric field of each particle by Liennard-Wiechert potentials:
\begin{equation}
	\label{eq13}
	\vec E_{\rm rad}(\vec r, t)  = \frac{q}{4 \pi \varepsilon_0} \frac{1}{\xi^3} \left[ \vec \rho \left( 1 - \vec \beta^2 + \vec D \vec \alpha \right) - \vec \alpha \left( \vec D \vec \rho \right) \right].
\end{equation}
where $\vec R$, $\vec \beta = d_{c t}\vec R$ and $\vec \alpha = d_{c t}\vec \beta$ are position, velocity and acceleration of a particle at the retarded time $t_{\rm ret} = t - \frac{1}{c} | \vec D |$. The other symbols are defined as $\vec D = \vec r - \vec R(t_{\rm ret})$, $\xi = | \vec D | - \vec D \vec \beta$, and $\vec \rho = \vec D - | \vec D | \vec \beta$. Our algorithm tracks the particles at the retarded time $t_{\rm ret}$ and linearly interpolates the fields to a given set of evaluation space-time-points to avoid solving the implicit equations.

\begin{figure}
	\includegraphics[width = 8.3 cm]{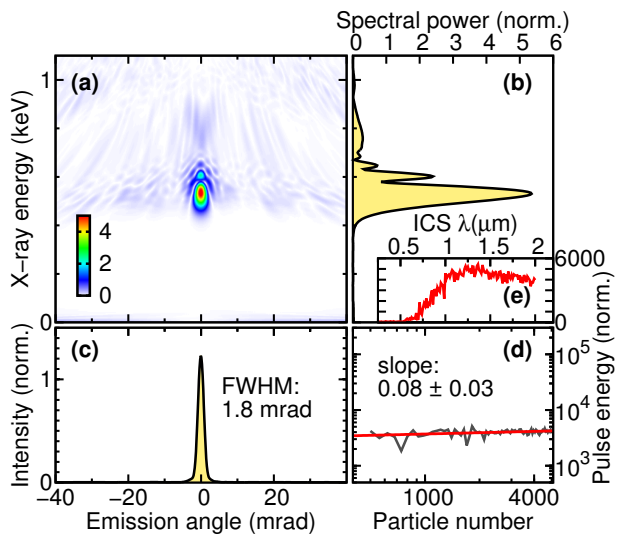}
	\caption{X-ray emission from the accelerated and compressed electron bunch colliding with an ICS laser pulse. (a) Integrated x-ray pulse energy as a function of emission angle and photon energy (normalized, color coded). Parameters: space charge for $Q = - 10~{\rm fCb}$, $N = 5000$ particles. (b) X-ray spectrum on-axis (normalized). (c) Integrated intensity vs. emission angle. (d) Coherent enhancement (see text). (e) Energy-scaling with ICS laser wavelength (see text). \label{fig04}}
\end{figure}
First, we set the offset $z_{\rm off} = 75 \, \mu {\rm m}$ such that the ICS pulse collides with the electron bunch at $ct = 65 \, \mu {\rm m}$ where it is compressed best. The bunch before acceleration is set to a smaller diameter of 0.5~$\mu$m and an lower energy spread of $\Delta W = 500 \, {\rm eV}$ transversally and $\Delta W = 100 \, {\rm eV}$ longitudinally. A large initial phase space volume, as used above, results in a corresponding reduction of coherent enhancement.

Fig.~\ref{fig04}~(a-c) shows the far-field x-ray spectra as a function of the emission angle for $N = 5000$ and $Q = - 10~{\rm fCb}$. A broadband incoherent background (light blue) is emitted into an angle of about 100~mrad with a chirp from 0.7~keV to 0.4~keV due to the lower acceleration of off-axis electrons. A spectrally and spatially narrow peak emanates from the coherent superposition of the radiation fields of all particles. The (FWHM) opening angle of 1.8~mrad resembles the Fourier-transform of the transverse bunch shape. A spectral width of about 80~eV at a center energy of 0.53~keV render these photons interesting for applications. Depending on the available acceleration laser, the concept is easily adapted to different x-ray ranges due to the favourable scaling of both, compression and ICS energy with $\gamma_f^2$ (see. Eq.~\ref{eq02}).

The degree of coherence is estimated from the scaling of the integrated squared fields with the number $N$ of macro-particles. Keeping the total charge $Q$ fixed, incoherent radiation yields a scaling as $N^{-1}$ whereas perfectly coherent emission is independent of $N$ as realized in our simulation (see Fig.~\ref{fig04}~(d)). At realistic electron numbers of $N=10^5$, the incoherent background becomes negligible. Increasing the electron count $N \propto Q$, the x-ray power grows as $N^2$. Even with a low total charge of 10~fCb, as choosen above, we estimate a x-ray pulse energy of 7~pJ (i.e. an ICS efficiency of $10^{-4}$). Due to relaxing requirements on the bunch, the pulse energy grows with the ICS laser wavelength $\lambda$, as shown in Fig.~\ref{fig04}~(e).

In conclusion, we have shown a scheme to achieve 15-as-short low emittance relativistic electron bunches by direct laser acceleration using low power lasers. We superimpose the acceleration laser and a counterpropagating inverse Compton scattering laser to create coherently enhanced x-ray pulses. 50-as x-ray bursts at 0.53~keV photon energy are predicted for realistic laser parameters. Our scheme might be employed as a low-emittance attosecond electron and x-ray source for ultrafast studies and for seeding of free electron lasers.

\begin{acknowledgments}
This work is supported by DARPA AXIS Program under grant N66001-11-1-4192. We thank W. S. Graves and L. J. Wong for helpful discussions and A. S. acknowledges funding by the Alexander von Humboldt Foundation.
\end{acknowledgments}

\bibliography{Bunch_Compression}

\end{document}